\documentclass[12pt]{article}
\usepackage{epsf}
\usepackage{epsfig}
\usepackage{wrapfig}
\usepackage{amsmath}
\usepackage{amssymb}
\usepackage{eufrak}
\usepackage{graphicx}
\newcommand{\link}{\mbox{\begin{picture}(4.15,10)
\put(0,2.4){\circle*{2}}                     \put(0,2.75){\line(1,0){12}}
\put(12.3,2.4){\circle*{2}}\mbox{  }  \end{picture}  }  }
\newcommand{\linkdual}{\;\link^{\!\!\!\!\!\!\!\!dual}}

\begin{document}

\begin{flushright}
УДК 539.12: 539.171 \end{flushright}
\vskip 15mm

\begin{center}
\section*{Non-locality requires fine tuning and multi-degenerate vacua}

\vskip 5mm
\underline{D.L. Bennett}$^{1}$ and H.B. Nielsen$^{2}$
\vskip 5mm

{\small
(1) {\it Brookes Institute for Advanced Studies, B\o gevej 6, 2900 Hellerup, Denmark} 
\\
(2) {\it The Niels Bohr Institute, Blegdamsvej 17, 2100 Copenhagen \O , Denmark}
\\
}
\end{center}
\vskip 5mm

\begin{abstract}
Our Multiple Point Principle (MPP) states that the realized values 
for e.g. the parameters of the standard model correspond to having 
a maximally degenerate vacuum. In the original appearence of MPP 
the gauge coupling values were predicted to within experimental 
uncertainties. A mechanism for fine-tuning follows in a natural 
way from the MPP. Using the cosmological constant as a example, 
we attempt to justify the assertion that at least a mild form of 
non-locality is inherent to fine-tuning. This mild form - namely 
an interaction between pairs of spacetime points that is identical 
for all pairs regardless of spacetime separation - is insured by 
requiring non-local action contributions to be reparametrization 
invariant. However, even this form of non-locality potentially 
harbours time-machine-like paradoxes. These are seemingly avoided
by the MPP fine-tuning mechanism. A (favorable)comparison of the 
results of MPP in the original lattice gauge theory context with 
a new implementation with monopoles that uses MPP at the transition 
to a monopole condensate phase is also described. 
\end{abstract}
\vskip 8mm

\subsection*{1. Introduction} 

The Multiple Point Principle (MPP) states that fundamental physical
parameters assume values that correspond to having a maximal number of
different coexisting ``phases'' in the vacuum.   
There is phenomenological evidence 
suggesting that some or all of the about 20 parameters in the Standard
Model (SM) that are not predicted within the framework of the SM correspond
to the MPP values of these parameters\cite{bennett1,bennett2,bennett3}. 
The MPP also provides a natural way to understand how fine-tuning comes
about in a manner analogous to the way that
the coexistence of the ice and liquid phases of water is required for a finite 
range of energies for the system all of which result in the fine-tuning
of the temperature to $0^oC$ (at 1 atm.). 

A cursory examination of the relation between the bare and dressed
cosmological constant that is maintained dynamically seems to demand at 
least a mild form of non-locality which in turn makes us vulnerable to
``matricide-like'' paradoxes. Postulating MPP as a law of Nature seems to  
avoid paradoxes and at the same time
provides an eloquent mechanism of fine-tuning standard model parameters to
experimentally observed values.

In this contribution, we outline a new way of applying MPP  by arguing
that new physics in the form of a fundamental (ontological) lattice
(and the accompanying monopoles) 
would necessarily appear at a scale for which an infinite  monopole self
coupling gives an upper limit (by the Triviality Theorem). 
The monopoles inherent to this lattice undergo a phase transition
to and from a monopole condensate phase. By arguing that letting  $\lambda$ run
to infinity would set the (upper limit for the) fundamental cutoff scale
and using that the gauge couplings approach constant values for ``large'' 
enough values of the 
self-coupling $\lambda$, we get predictions for the gauge couplings
by letting a curve that shares a point with a monopole transition curve
at small (negative) $\lambda$ values run to infinite $\lambda$. These
predictions are 
compared with the original lattice gauge theory implementation of MPP.

\subsection*{2. Non-locality seems inherent to the problem of fine tuning}

In essence, solving the problem of fine tuning means finding a way to
render couplings (or other intensive quantities) dynamical. This invariably 
leads to non-locality - at least in a mild sense - in a way exemplified
by the fine tuning problem for the dressed cosmological constant
\cite{bennett4}. If a
coupling, e.g., the cosmological constant, is dynamical,  locality dictates 
that the value of the coupling at some spacetime point can directly depend 
on that spacetime point and indirectly on other spacetime points at earlier 
times but certainly not on the future.

But if the bare cosmological constant immediately following
the big bang is to already have its value fine tuned very exactly to the value 
that makes the dressed cosmological
constant as small as that suggested phenomenologically, we have a
problem with locality in the following sense: in order that the  bare
cosmological constant be relateable to the value of the dressed cosmological 
constant, the details of the dressed
cosmological constant 
that will evolve in the future must be
known at the time of big bang. So a strict principle of locality is not allowed
if we want to have a dynamically maintained bare coupling and renormalization 
group corrections of a quantum field theory with a well-defined vacuum.

An allowable resolution would be to allow a mild class of 
non locality consisting of an interaction that is
the same between any pair of points in spacetime independent of the distance 
between these points. 
It would be difficult to see that such an interaction is non-local.
Instead, such spacetime
omnipresent fields - a sort of background that is forever everywhere the 
same - would likely be interpreted simply as constants of Nature.

The reparametrization invariance of general relativity
implies this symmetry (i.e., the same interaction 
between any pair of spacetime points). 
Leting $\phi(x)$ stand for all the fields (and derivatives of same) 
of the theory, we use the fact that integrals (extensive quantities) of the form
$I_{f_j}[\phi(x)]\stackrel{def}{=}\int dx^4\sqrt{g(x)}f_j(\phi(x))$ as well as any
function of such integrals are  reparametrization invariant. Here the $f_j(\phi)$ 
are typically Lagrange densities.  We get a
reparametrization invariant but non-local action $\hat{S}_{nl}$ by taking a 
non-linear function of the funtionals $I_{f_j}[\phi(x)]$ as $\hat{S}_{nl}$. 

Each extensive quantity $I_{f_j}[\phi(x)]$ has a fixed value - call
it $I_{f_j \; fixed}$ (fixed in the
sense of being a Law of Nature) for each imaginable 
Feynman path integral history of the Universe as it evolves from Big Bang at time
$t_{BB}$ 
to Big Crunch at time $t_{BC}$. The set $\{I_{fixed\;f_j} \}$ of allowed values for 
the $I_{f_j}$
can be implemented as a $\delta$-function in the functional integration measure that
defines our reparametrization invariant non-linear (and therefore non-local)
action $\hat{S}_{nl}$: 
$\exp(S_{nl}(\{I_{f_j}\})=\prod_j\delta(I_{f_j}[\phi]-I_{fixed\;f_j})$.

\subsection*{3. The Multiple Point Principle (MPP)}

The first appearence\cite{bennett1} of the MPP was in connection with 
predictions of the values of the non-Abelian gauge couplings.
This was done in the the context of lattice gauge theory
using our so-called family replicated gauge 
group\cite{bennett4a,bennett5} $G_{FRGG}$ 
(also sometimes referred to as
$G_{Anti-GUT}$) which consists of the 3-fold replication
of the Standard Model Group (SMG): 
$G_{FRGG}=SMG \otimes SMG \otimes SMG \stackrel{def}{=}
SMG^3$ (in the extended version: $(SMG\times U(1))^3$ )
having one $SMG$ factor for each generation of fermions
and gauge bosons. We postulate that $G_{FRGG}$ is broken to
the diagonal subgroup (i.e., the usual SMG) at roughly the Planck scale.

In the original context of predicting the standard model gauge couplings, 
MPP asserts that the Planck scale values of the
standard model gauge group couplings coincide with the multiple point, i.e.,  
the point or (hyper)surface that lies in the boundary separating the maximum 
number of phases in 
the action parameter space corresponding to the gauge group
$G_{FRGG}$. The (Planck scale) 
predictions for the gauge couplings are subsequently  
identified with the
parameter values at the point in the action parameter space for the diagonal 
subgroup of $G_{FRGG}$ that is
inherited from the multiple point for $G_{FRGG}$ after the Planck
scale breakdown of the latter.

The phases to which we refer are usually dismissed as lattice artifacts
(e.g., a Higgsed phase, a confined or Coulomb-like phase).
Such phases have been studied extensively in the literature for 
gauge groups such as $U(1)$, $SU(2)$ and $SU(3)$). One typically finds first 
order phase transitions
between confined and Coulomb-like phases at critical values of the action
parameters.   

We suggest that these lattices phases correspond to phases
that may be inherent to any regulator. As a
regulator in some form (be it a lattice, strings or whatever) is
needed for the consistency of any quantum field theory, it is consistent to 
assume the existence of a fundamental regulator.
The ``artifact'' phases that arise in a theory with such a  
regulator (that we have initially chosen to implement as
a fundamental lattice) are accordngly taken as ontological phases that have
physical significance at the scale of the fundamental regulator (e.g.,
lattice). The assumption of an ontological fundamental regulator implies
the existence of monopoles in terms of which the regulator induced phase
can also be studied\cite{bennett5}. 

Finding the multiple point in an action parameter space corresponding to
the gauge group $G_{RGG}$ is more complicated than for groups such as $U(1)$,
$SU(2)$ or $SU(3)$ say. The boundaries between phases in the action parameter
space (i.e., the phase diagram) must be sought in a high dimensional parameter 
space essentially because $G_{Anti-GUT}$ being a non-simple group has many
subgroups and invariant subgroups.

In fact there is a distinct phase for each subgroup pair $(K,H)$ 
where $K$ is a 
subgroup and $H$ is an invariant subgroup such that 
$H \triangleleft K \subseteq G_{FRGG}$. An
element $U \in G_{FRGG}$ can be parameterized as $U=U(g,k,h)$
where the Higgsed (gauge) degrees of freedom 
are elements $g$ of the homogeneous space $G_{FRGG}/K$. 
The (un-Higgsed) Coulomb-like and confined degrees of freedom
are respectively the elements $k$ of the factor group $K/H$ and the 
elements $h \in H$. 

\subsection*{4. The History of the Universe as a Fine-tuner}

MPP functions as a fine-tuning mechanism when the extensive quantities
$I_{fixed\;f_j}$ introduced in Section 2 happen to have values that can 
only be realized in a universe
having two (or more) coexisting phases the transition between which
is first order in which case the intensive quantity (typically a 
coupling) conjugate to $I_{fixed\;f_j}$ is fine-tuned. It may be useful to
consider a familiar analogy: for a system consisting of $H_2O$ there is
a whole range of energies (corresponding to the heat of melting) for which
the system is forced to be realized as coexisting ice and liquid phases in
which case the energy-conjugate intensive parameter i.e., temperature, 
is fine-tuned
(e.g., to $0^oC$ for a presure of 1 atm.). A better implementation  
of MPP would be the triple point of water where for a finite range
of energies and volumes  three phases meet and 3-1=2 
intensive parameters (the temprature and presure) are fine-tuned to the
triple point values. 

In 4-space, one generic possibility for having coexistent phases would be to 
have a phase with $\phi_{us}$ in an early epoch including say the universe 
as we know it and a phase with $\phi_{other}$ in a later epoch:  
\begin{equation}
I_{fixed\;f_j}=
f_j(\phi_{us})(t_{ignit}-t_{BB})V_3+ f_j(\phi_{other})(t_{BC}-t_{ignit})V_3\;\;\;
\small{(\mbox{$V_3$ is the 3-volume of the universe})}
\label{coexist}\end{equation}
where $t_{ignit}$ is the ``ignition'' time (in the future) at which there is a 
first
order phase transition from the vacuum at $\phi_{us}$ to the later
vacuum at $\phi_{other}$. 
The value of the ``coupling constant'' conjugate to $I_{fixed\;f_j}$ gets fine 
tuned (unavoidably by assumption
of the coexistence of the two phases separated by a first
order transition) by a mechanism that also depends on a phase that will 
first be realized in the future (at $t_{ignit}$). Such a mechanism is  
non-local. Note in particular that the right hand side of Eqn.~\ref{coexist} 
depends on $t_{ignit}$.  

We want to formally define a ``coupling constant'' 
conjugate to some extensive quantity $I_{fixed\;f_j}$.  
Restrict the non-local 
action $\hat{S}_{nl}=\hat{S}_{nl}(\{I_j[\phi(x)]\})$ to being a non-local
potential $V_{nl}$ that is a function of (not necessarily independent) 
functionals: $V_{nl}
\stackrel{def}{=} V_{nl}(I_{f_i}[\phi],I_{f_j}[\phi],\cdots)$. 
Define an effective potential $V_{eff}$ such that
\begin{equation}
\frac{\partial V_{eff}(\phi(x))}{\partial \phi(x)} \stackrel{def}{=}
\frac{\delta V_{nl}(\{I_{f_j}[\phi]\})}{\delta \phi(x)}
\left|_{near\;\;min.} \right.=
\sum_i\left(\frac{\partial V_{nl}(\{I_{f_j}\})}{\partial I_{f_i}}\frac{\delta
I_{f_i}[\phi]}
{\delta \phi(x)}\right)\left|_{near\;\;min.} \right. \label{eq11}
\end{equation}
\[ =\sum_i\frac{\partial V_{nl}(\{I_{f_j}\})}{\partial I_{f_i}}
\left|_{near\;\;min.} \right. f_i^{\prime}(\phi(x))  \]
The subscript ``near min'' 
denotes  the approximate ground state of the whole universe, up to deviations
of $\phi(x)$ from  its vacuum value (or vacuum values for a multi-phase vacuum)
by any amount in relatively small spacetime regions.
The solution to Eq.~(\ref{eq11}) is
\begin{equation}
V_{eff}(\phi)=\sum_i \frac{\partial V_{nl}(\{I_{f_j}\})}{\partial 
I_{f_i}}f_i(\phi)
\label{eq12} \end{equation}
We identify 
$\frac{\partial V_{nl}(\{I_{f_j}\})}{\partial I_{f_i}}$ as intensive quantities
conjugate to the $I_{f_i}$.

Consider now the effective potential
(\ref{eq12}) in the special case that
$V_{nl}(\{I_{f_j}\})=V_{nl}(I_2,I_4)\stackrel{def}{=}
V_{nl}(\int d^4x\sqrt{g(x)}\phi^2(x),\int d^4y\sqrt{g(y)}\phi^4(y))$
in which case, (\ref{eq12}) becomes
\begin{equation} V_{eff}=\frac{\partial V_{nl}(I_2,I_4)}{\partial 
I_2}\phi^2(x)+
\frac{\partial V_{nl}(I_2,I_4)}{\partial I_4}\phi^4(x)\stackrel{def}{=}
\frac{1}{2}m^2_{Higgs}\phi^2(x)+
\frac{1}{4}\lambda\phi^4(x) \label{msqh}\end{equation}
where the right hand side of this equation, which also defines 
the (intensive) couplings $m^2_{Higgs}$ and
$\lambda$, is recognised as a prototype scalar potential
at the tree level.
Of course the
form of $V_{nl}$ is, at least {\em a priori}, completely unknown to us,
so - for example - the coupling
constant $m^2_{Higgs}$ cannot be calculated from Eqn.~\ref{msqh}.
The potential of Eqn.~\ref{msqh}
with $m_{Higgs}^2<0$ has an asymmetric minimun at, say, the value $\phi_{us}$ 
resulting in 
spontaneous symmetry breakdown 
in the familiar way. 

Actually we want to consider the  potential $V_{eff}$ having the two relative 
minima 
$\phi_{us}$ and $\phi_{other}$ - both at nonvanishing values of $\phi$ - 
alluded to at the beginning of this section.
The second minimum comes about at a value $\phi_{other}>\phi_{us}$ when 
radiative corrections to (\ref{msqh}) are taken into 
account and the top quark mass is not too 
large\cite{bennett3,bennett6,bennett7}.
Which of these vacua - 
the one at $\phi_{us}$ or $\phi_{other}$ - 
would be the stable one in this 
two-minima Standard Model effective Higgs field potential depends on
the value of $m^2_{Higgs}$. 
Since $I_2$ and $I_4$ are  functions of $t_{ignit}$ (as seen from 
Eqn.~\ref{coexist} with $f_j=\phi^2$ or $\phi^4$),
$m^2_{Higgs}\stackrel{def}{=}
\frac{\partial V_{nl}(\{I_2,I_4\})}{\partial I_2}$ is also a function of
$t_{ignit}$.

\begin{figure}
\vspace{-2cm}
\centerline{\epsfxsize=\textwidth \epsfbox{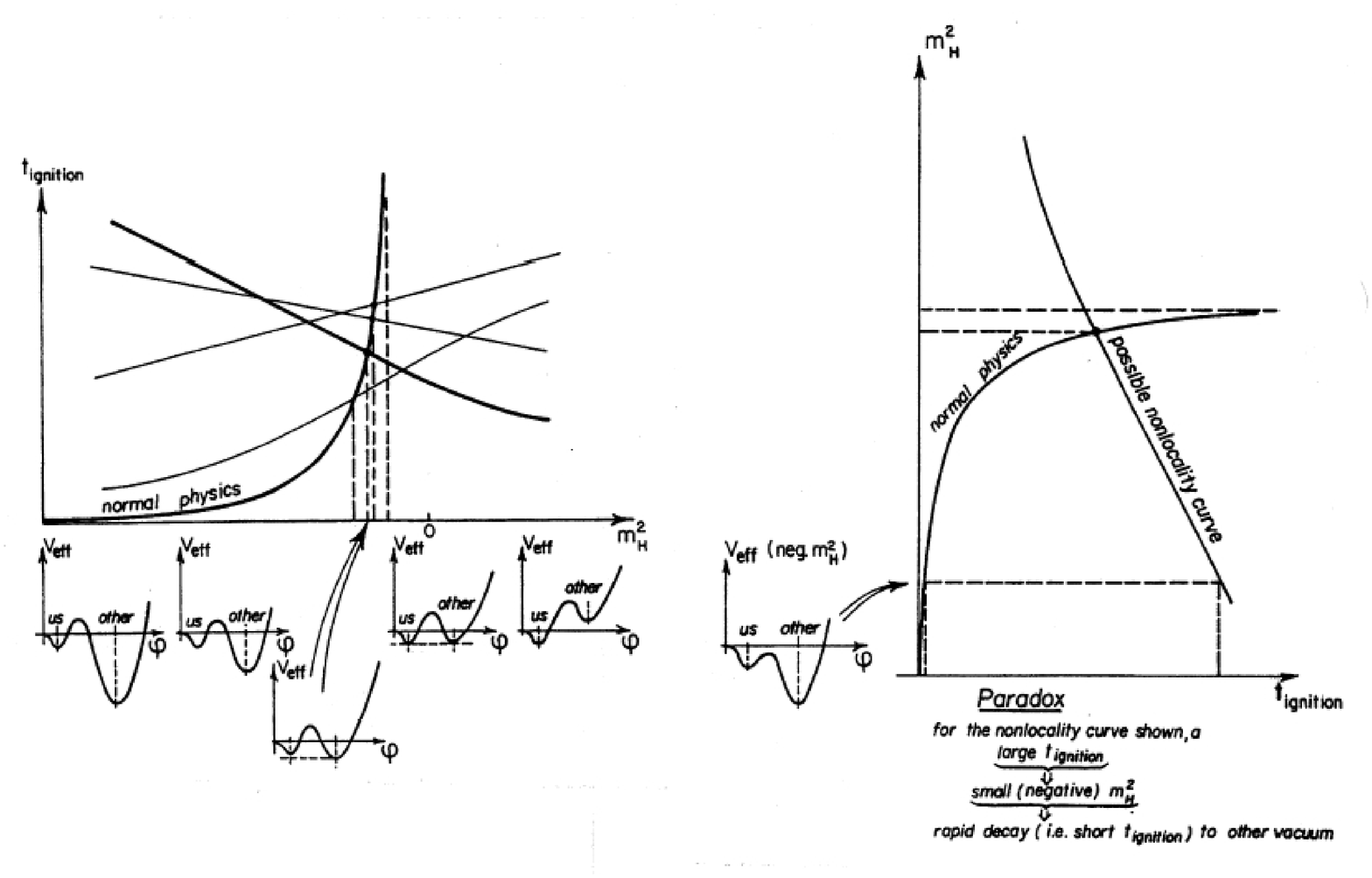}}
\vspace{-.5cm}
\caption[figurenonloc]{\label{fignonloc} (Left) The development of the double 
well
potential and $m_{Higgs}$ as a function of $t_{ignit}$. Note that
all the more or less randomly drawn non-locality curves intersect the
``normal physics'' curve
near where the vacua are degenerate (i.e., the MPP solution).}
\vspace{-.2cm}
\caption[figparadox]{\label{paradox} (Right) Many non-locality curves 
could lead to
paradoxes similar to the ``matricide'' paradox. Such paradoxes are avoided if
the value of $m_{Higgs}$ is fine-tuned to the multiple point critical value.
This corresponds to the intersection of the ``normal physics'' curve with the 
``possible nonlocality'' curve.}
\vspace{-.3cm}
\end{figure}
Let us first use  ``normal physics'' to see how the relative depths of the
two minima of the double well are related to $m^2_{Higgs}$ and to
$t_{ignit}$. It can be deduced from\cite{bennett7} that a large
negative value of $m^2_{Higgs}$ corresponds to the relative minimum 
$V_{eff}(\phi_{other})$ being deeper than $V_{eff}(\phi_{us})$
(in which by assumption the Universe starts off following Big Bang)
than for less negative values
of $m^2_{Higgs}$ (see Fig.~\ref{fignonloc}). It can also be argued quite
plausibly that a minimum in $V_{eff}$ at $\phi_{other}$ much deeper
than that at $\phi_{us}$ would correspond to an
early (small) $t_{ignit}$ inasmuch as the ``false'' vacuum at
$\phi_{us}$ would be very unstable. However, as the value
of the potential at $\phi_{other}$ approaches that at
$\phi_{us}$, $t_{ignit}$ becomes longer and longer
and approaches infinity as the values of $V_{eff}$ at
$\phi_{us}$ and $\phi_{other}$
become
the same. 
The development of the double well potential and
$m^2_{Higgs}$ as a function of $t_{ignit}$ is illustrated in 
Fig.~\ref{fignonloc}.
Note that the larger the difference
$|\phi_{other}-\phi_{us}|$
the more the realization of say $I_{fixed\;2}$ will in general depend on 
$t_{ignit}$. 
If $\phi_{us}=\phi_{other}$, 
$t_{ignit}$ plays no role in realizing e.g. $I_{fixed\;2}$ and the value
of $m^2_{Higgs}$ becomes independent of $t_{ignit}$.

\subsection*{5. Avoiding paradoxes arising from non-locality}

In general the presence of non-locality leads to paradoxes. While
the form that the non-local action (or potential $V_{nl}$ in this discussion)
is unknown to us, we make the 4 generically representative guesses 
portrayed as the 4 non-locality curves 
in Fig.~\ref{fignonloc}. 
In particular, non-locality curves having a negative slope as a function
of $t_{ignit}$ lead to paradoxes in the following manner.
Consider the non-locality curve in Fig.~\ref{fignonloc} drawn with bold line
that is redrawn in a rotated position in Fig.~\ref{paradox}. Let us make the
assumption that  $t_{ignit}$ is large and see that this leads to a
contradiction. Assuming that $t_{ignit}$ is large, it is seen from
the non-locality function in Fig. \ref{paradox} 
(call it $m^2_{Higgs\;nl}(t_{ignit})$ to distinguish it from the 
``normal physics'' $m_{Higgs}^2(t_{ignit})$) that this
implies that the ``normal physics'' $m^2_{Higgs}$ has a large negative value.
But a large negative value of $m^2_{Higgs}$ corresponds in ``normal physics'' 
to a (false) vacuum at $\phi_{us}$ that is very
unstable and therefore to a very short $t_{ignit}$ corresponding to a rapid
decay to the stable vacuum at $\phi_{other}$. So the
paradox
appears: the assumption of a {\em large} $t_{ignit}$ implies a
{\em small} $t_{ignit}$. This happens because in general 
$m^2_{Higgs}(t_{ignit})\neq m^2_{Higgs\;nl}(t_{ignit})$ and is akin 
to the ``matricide''
paradox encountered for example when dealing with ``time machines''.
It has been suggested \cite{bennett8,bennett9,bennett10} that Nature avoids such paradoxes by
choosing a very clever solution in situations where these paradoxes lure.

In the case of the paradoxes that can come about due to non-locality of the
type considered here, a clever solution that avoids paradoxes is available to 
Nature in the form of the Multiple Point Principle (MPP).
The MPP solution corresponds
to the intersection of the ``normal physics'' curve and the ``non-locality
curve'' in Fig.~\ref{paradox}. 
because here the vacua at $\phi_{us}$ and $\phi_{other}$ are (essentially) 
degenerate. But at this intersection point, $m_{Higgs}^2(t_{ignit})=
m^2_{Higgs\;nl}(t_{ignit})$
so the paradox is avoided. So the paradox is avoided at the multiple point.
But at the multiple point, an intensive
parameter has its value fine-tuned for a wide range of values of the 
conjugate extensive quantity.
Fine-tuning can therefore be understood as a
consequence of Nature's way of avoiding paradoxes that can come about due to
non-locality.

\subsection*{6. Determining MPP gauge couplings at transition to monopole
condensate}    

Until now we have talked about the determination of gauge couplings 
by finding the point (or surface) - the multiple point (surface) - where the 
maximum number of ``lattice artifact''
phases come together in the action parameter space of a lattice 
gauge theory. Here we briefly sketch an alternative that potentially replaces 
the assumption of an ontological lattice by instead 
producing the different phases 
using 
the assumption of monopoles. Monopoles can cause different 
phases by
condensing or not. 
Then the MPP prediction is that the couplings realized in Nature
are the values at the transition between a Coulomb-like phase and
a monopole condensate. Because it is not important whether the monopoles
are lattice artifact monopoles or fundamental monopoles or whatever,
this way of determining the MPP gauge couplings offer the possibility
of exonerating the original method that considers ``lattice artifact''
phases. 

A simple effective dynamical description of confinement in a pure $U(1)$
lattice gauge theory is the dual Abelian Higgs model of scalar 
monopoles\cite{bennett11} 
We shall shortly see that the dualized $U(1)$
lattice theory can be rewritten as an $\mathbf{R}$ lattice gauge theory on 
the dual
lattice with a ``non-linear'' Higgs field. In earlier work\cite{bennett12}
the formulation dual to
the renormalization group improved Coleman-Weinberg effective potential
(at tree level this is just Eqn. \ref{msqh}) was considered in the two 
loop approximation. In this work the transition to a monopole condensate
was found along a phase transition curve in $(\lambda, g^2)$ space situated 
in the region with 
negative $\lambda$ of the order of -10. Here $\lambda$ and $g^2$ are
running couplings for which one would think that the renormalization point 
$\mu$ should be taken to be of the order of the Higgs monopole mass or the VEV
of the monopole field in the condensed phase. The negative $\lambda$ is not
alarming for the existence of the phase transition even if one requires a 
bottom for the Hamiltonian because by running to a higher renormalization point
(identified with field strength) the self-coupling can run positive.  
In the earlier work\cite{bennett12} the MPP philosophy was to take the 
actual coupling
directly related to the couplings on the phase transition curve just 
mentioned. 
In the present paper we want to consider MPP applied to bare couplings. 
If bare couplings are considered, one must declare the scale of the bare 
coupling. To avoid a negative $\lambda$, one should at least choose the 
ontological cutoff scale high enough that $\lambda$ is positive.
On the other hand 
we know from the Triviality Theorem that new physics (e.g., a
fundamental lattice) must show up at the latest at the energy where the 
monopole self-coupling 
$\lambda$ becomes infinite due to renormalization group running.

So we have an allowed interval for the scale of the fundamental lattice:
it must be greater than that for which $\lambda_{running}$ becomes positive
but less than the scale at which $\lambda$ runs to infinity.

Now it turns out that as $\lambda$ runs to ``large'' values, the dependence
of $g^2$ on $\lambda$ becomes very weak. So by running the 
$g^2(\lambda)$ curve {\it from} the one point that it has in
common with the above-mentioned phase transtion
curve along which there is a Coulomb - condensate phase transition 
{\it to}  ``large'' 
enough $\lambda$
values, we get a good candidate for the running $g^2$ value at the scale of the
fundamental regulator (lattice).

So what we now need is a justification for assuming that $\lambda$ is ``large''
enough. There are a couple of justification candidates.
One approach would be to make the assumption that the approximation 
of a continuum
spacetime is good all the way up to the regularization scale given by the 
Triviality Theorem bound. Since this bound is determined by 
$\lambda \stackrel{running}{\rightarrow} \infty$ we have justified 
the assumption of large $\lambda$. The second approach to justifying the 
assumption of large $\lambda$ entails the assumption of an ontological
lattice and a little calculation.

The calculation would start by using dualization to rewrite a
$U(1)$ lattice gauge theory into a theory with the gauge group $\mathbf{Z}$ 
(under addition) on the dual lattice links $\linkdual$. 
Next we endeavor to get this 
integer
gauge group from the group of real numbers $\mathbf{R}$ by
using a  weighting which for each dual link $\linkdual$
is a sum of infinitely many
$\delta$-functions in the gauge group $\mathbf{R}$ minus an integer $n$ over 
which we sum:

\[ \prod_{\linkdual}\sum_{n\in \mathbf{Z}}
\delta(\theta_{\mathbf R}(\linkdual)-2\pi n). \]

By putting in the $\delta$-function factor, we essentially go from
a real number gauge group to an integer gauge group. Now the trick
is to get this series of $\delta$-function weightings for each dual link by 
formally introducing a series of ``non-linear'' Higgs fields
i.e., a scaler field on each site that has the complex unit circle as its
target space. By choosing the unitary gauge we can take the value unity
for this non-linear gauge field. Then the lowest order action contribution
(from the kinetic term of the Higgs field) will be 
$\beta_{\mathbf{R}}cos(\theta_{\mathbf{R}}(\linkdual))$
for each dual link (summed over all dual links).
If $\beta_{\mathbf{R}}$ goes to infinity, the exponentiated action for each link 
will be a weight of the form we want (i.e., a sum of infinitely many
$\delta$-functions).

\subsection*{7. Conclusion}

We attempt to justify the assertion that fine-tuning in Nature 
seems to imply a fundamental form of non-local interaction. 
This could be manifested in a phenomenologically acceptable form as everywhere 
in spacetime 
identical interactions between any pair of spacetime points. This would be 
implemented by requiring the non-local action to be diffeomorphism invariant. 

Next we put forth our multiple point principle\cite{bennett1,bennett2} 
which states that coupling
parameters in the Standard Model tend to assume values that correspond to
the values of action parameters lying at the junction of a maximum number of
regulator induced phases (e.g., so-called ``lattice artifact phases'') 
separated
from one another in action parameter space by first order transitions. 
The action. 
which of course is defined on a gauge group 
(e.g., the non-simple SM gauge group) governs fluctuation patterns
along the various subgroup combinations $(K,H)$ with $H \lhd  K \subseteq G$
that characterize the phases that come together at the multiple point.

We then consider extensive quantities that are functions 
of functionals $I_{f_j}[\phi(x)]$ that are essentially Feymann path histories
of the Universe for functions $f_j(\phi)$
of the fields $\phi (x)$ and derivatives of these fields. 
We then think of the generic situation in which
these extensive quantities can happen to be 
fixed 
at values  that require the universe to be realized as two or more
coexisting phases\cite{bennett4}. 
We draw on the analogy to the forced coexistence of ice 
and liquid water that occurs for a whole range of possible total energies
because of the finite heat of melting (first order phase transition).
With our multiple point principle, the intensive quantities (couplings)
conjugate to extensive quantities fixed in this way  become fine-tuned in a 
manner analogous to the fine
tuning of temperature to $0^oC$ (at 1 atm.) when the total energy of a system 
of $H_2O$ is such that the system can only be realized as coexisting ice and 
liquid phases.
    
One generic way of having  coexisting phases in a quantum field
theory in 3+1 dimensions would be to have different phases in
different epochs of the lifetime of the Universe with phase transitions
occuring at various times in the course of the lifetime of the Universe.
If the transitions were first order, one would have fine-tuning of
(intensive) couplings conjugate to extensive quantitity values that can only be
realized by having  coexisting (i.e., more than one) phases. 
But such a fine-tuning would 
involve non-locality: the fine-tuned values of coupling constants would 
depend on future phase transitions into phases that do not even exist at
the time such couplings are fine-tuned.

Even non-locality of this sort (i.e., non-localy manifested as
a diffeomorphism invariant contribution to the action) can lead to paradoxes 
of the
``matricide paradox'' type.  We argue that such paradoxes are avoided when 
Nature chooses the multiple point principle solution to the problem of 
finetuning\cite{bennett4}.

\begin{table}
\caption{
Table of values of $\alpha^{-1}(\mu_{Pl.})$. Recall that these values 
are those for {\it each} of the three $U(1)$s, {\it each} of the three 
$SU(2)$s and {\it each} of the three $SU(3)$s in our family replicated gauge 
group 
$G_{FRGG}$ because $G_{FRGG}$ is the 3-fold 
Cartesian product of the usual standard model group (SMG). Following the 
Planck scale breakdown of $G_{FRGG}$ to the diagonal subroup (isomorphic to 
the usual SMG)
the $\alpha^{-1}(\mu_{Pl.})$ values for the 
non-Abelian subgroups get multiplied
by a factor 3 whereas $\alpha^{-1}_{U(1)}(\mu_{Pl.})$ gets enhanced by a
factor somewhat greater than six ($ \approx 6.5 $) for reasons having to do 
with
the three $U(1)$s of $G_{FRGG}$ being Abelian.\cite{bennett2} For comparison,
the last row gives experimental values of $\alpha^{-1}(\mu_{Pl})$ that have 
been extrapolated to Planck scale using the renormalization 
group (with minimal standard model) and
subseqently divided by factors 3 and 6.5 in respectively the non-Abelian and
Abelian cases.} \vspace{.5cm}
\begin{tabular}{|l||l|l|l|} \hline
  & U(1) & SU(2)/{\bf Z}$_2$ & SU(3)/{\bf Z}$_3$ \\ \hline
Naive continuum limit & 12.4 &  21.7 &  26.7 \\ \hline
Parisi Improved & 8.25  & 15.4 & 16.3 \\ \hline
Monopole 1-loop & 7.20 & 15.0 & 18.2  \\ \hline
Experimental values (Planck scale)\footnote{where?}& 8.22 & 16.4  & 17.9 \\ 
\hline
\end{tabular} 
\end{table}

The first formulation of the MPP arose in our predictions of the SM gauge
coupling constants\cite{bennett1,bennett2}. This was done in the context of 
lattice gauge theory
using our family replicated gauge group  $SMG \otimes SMG \otimes SMG$ having
one SMG for each family of fermions (and bosons!)\cite{bennett4a,bennett5}. 
This is broken to the
diagonal subgroup (which is isomorphic to the usual SMG) at roughly the Planck
scale. The predictions for the ``Naive continuum limit'' and ``Parisi Improved''
values of $\alpha^{-1}(\mu_{Pl})$ (see table) rely on
on the assumption that what are usually regarded a ''latice artifact'' are in fact
ontological. This assumption is avoided in  recent work that uses a monopole
technology to make MPP predictions of gauge couplings. 
These is in rather good agreement of the ``Monopole 1-loop''
with the original predictions (see table of $\alpha^{-1}(\mu_{Pl.})$ values)

We use the Triviality Theorem in support of our requirement that the monopole
(on the dual lattice) attain an infinite self-coupling at the lattice scale. 
This requirement excludes having a fixed point in the  running $\lambda$ value
before reaching the lattice scale because $\beta_{\lambda}=0$ at any scale under 
that of the lattice would stop the running before $\lambda$ becomes infinite.    

In the two loop approximation it turns out that zeros in $\beta_{\lambda}$ show
up at $ g^2 \approx 19$ or above. But if we trust the Triviality Theorem, we must 
assume that
going to three loops (or more) would remove the two loop (artifact) $\beta_{\lambda}$ 
zeros or that we already are at sufficiently strong $g^2,\lambda$ so that the optimal
order is one-loop and that going to two-loops introduces error. For now,
we use the one loop calculation in calculating the values in the table and
postpone a two loop or other attempts at a more accurate treatment.

\end{document}